# Agentic Workflow for Education: Concepts and Applications

**Yuan-Hao JIANG**[a,b†], **Yijie LU**[c†], **Ling DAI**[c,d†], **Jiatong WANG**[a,b], **Ruijia LI**[a,c] **& Bo JIANG**[a,b]*

[a]*Shanghai Institute of Artificial Intelligence for Education,*
*East China Normal University, Shanghai, China*
[b]*Lab of Artificial Intelligence for Education, East China Normal University, Shanghai, China*
[c]*Faculty of Education, East China Normal University, Shanghai, China*
[d]*National Institute of Education, Nanyang Technological University, Singapore*
*\* Corresponding author: bjiang@deit.ecnu.edu.cn*
*† These authors contributed equally to this work.*

**Abstract:** With the rapid advancement of Large Language Models (LLMs) and Artificial Intelligence (AI) agents, agentic workflows are showing transformative potential in education. This study introduces the Agentic Workflow for Education (AWE), a four-component model comprising self-reflection, tool invocation, task planning, and multi-agent collaboration. We distinguish AWE from traditional LLM-based linear interactions and propose a theoretical framework grounded in the von Neumann Multi-Agent System (MAS) architecture. Through a paradigm shift from static prompt–response systems to dynamic, nonlinear workflows, AWE enables scalable, personalized, and collaborative task execution. We further identify four core application domains: integrated learning environments, personalized AI-assisted learning, simulation-based experimentation, and data-driven decision-making. A case study on automated math test generation shows that AWE-generated items are statistically comparable to real exam questions ($P \geq 0.439$), validating the model's effectiveness. AWE offers a promising path toward reducing teacher workload, enhancing instructional quality, and enabling broader educational innovation.

## 1. Introduction

The emergence of Large Language Models (LLMs) has profoundly transformed the landscape of education (Dai, Jiang, et al., 2025). Their accessibility and affordability have facilitated a proliferation of practical applications across educational contexts (N. Singh et al., 2023). Agentic workflows powered by LLMs hold the potential to alleviate teachers' workloads by automating tedious and repetitive tasks (Andrew, 2024; Jiang,

Li, Wei, et al., 2024). However, current interactions with LLMs remain predominantly linear, wherein a user issues a prompt and the model generates a one-off response. This question-answering paradigm is often characterized as a non-agentic workflow (A. Singh et al., 2024). In such one-turn exchanges, LLMs lack opportunities for self-reflection or iterative refinement, thus functioning more as advanced "search engines" than as autonomous agents.

The advent of LLM-based Artificial Intelligence (AI) agents marks a significant shift from this static paradigm. Rather than treating LLMs as passive responders awaiting user input, agents imbue them with the ability to plan, reflect, and act autonomously—perceiving their environment, making decisions, and executing actions accordingly (Dai, Jin, et al., 2025; Jiang, Shi, et al., 2024). By incorporating self-reflection, tool invocation, task planning, and multi-agent collaboration, these agentic workflows exhibit the capacity to address complex, open-ended problems (Andrew, 2024). Their emergence not only revitalizes agent technologies but also opens up new avenues for the integration of AI into educational practices. The main contributions of this study are as follows:

- **Conceptual advancement:** We define the Agentic Workflow for Education (AWE) and distinguish it from traditional LLM-based interaction by establishing a four-component model—self-reflection, tool invocation, task planning, and multi-agent collaboration—as illustrated in Figure 1.
- **Framework proposal:** We extend the von Neumann Multi-Agent System (MAS) Framework to construct a theoretically grounded AWE structure, linking agent components with educational workflow stages and levels of technological maturity.
- **Application Pathways:** We analyze four core AWE applications—integrated learning environments, personalized learning, simulation-based scenarios, and data-informed decision-making—backed by empirical evidence.

## 2. Related Work

*2.1 From Workflows to Agentic Workflows*

An AI agent is a software system capable of responding to natural language inputs and performing tasks aligned with user needs (Gates, 2023). The concept of workflows can be traced back to the WorkFlo Business System (WBS), developed by FileNet in 1984, and to the specifications defined by the Workflow Management Coalition (WMC) in the 1990s (Workflow Management Coalition, 1995, 1996). Building on these foundational developments, a widely adopted model of agentic workflow reasoning comprises four dimensions: self-reflection and tool invocation provide the technical foundation, while task planning and multi-agent collaboration represent more advanced developmental directions (Andrew, 2024). This architectural approach improves processing efficiency in complex scenarios by enabling autonomous task execution and iterative optimization.

In educational contexts, agentic workflows exhibit strong domain adaptability. These workflows are constructed with AI agents as the core technology and workflow structures as the organizational framework, jointly forming a coherent system. For example, in automated essay grading, the workflow sequentially invokes four types of agents—data processing, text analysis, scoring, and feedback generation—enabling end-to-end automation from input to personalized output. Notably, the EXplanation-AugMented Training (EXAM) framework built on the Tongyi Qianwen LLM achieves a grammar error recognition accuracy of 69.76% under 15,000 training samples (Y. Li et al., 2024).

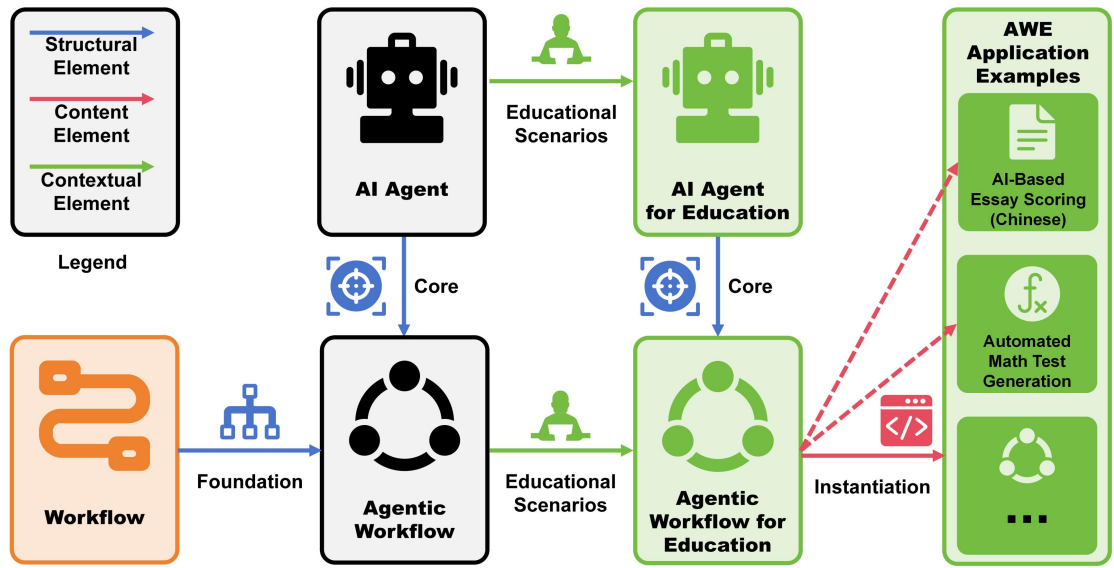

*Figure 1*. Conceptual framework of the agentic workflow for education

In another case, a collaborative system involving domain experts, question generation, automatic problem solving, and option generation agents showed no statistically significant differences from human-generated test items in contextual appropriateness ($P$ = 0.439) and option rationality ($P$ = 1.000). Furthermore, it significantly outperformed GPT-4 in option rationality ($P$ = 0.043) and question stem coherence ($P$ = 0.009) (R. Li et al., 2024). These findings suggest that agentic workflows contribute to improved educational outcomes through three mechanisms: (1) modular division of labor among agents ensures precision in task processing; (2) workflow self-optimization mechanisms support continuous improvement; and (3) collaborative agent behavior generates emergent intelligence effects.

## 2.2 New Paradigm: Agentic Workflows for Education

Large Language Models (LLMs) serve as the foundation of AI agents, enabling natural language interaction and supporting a wide range of educational tasks through integrated perception, decision-making, and execution. As shown in Figure 1, the evolution from general workflows to AWE progresses through layered integration—from basic automation to AI agent orchestration, ultimately forming domain-specific workflows shaped by structural, content-based, and contextual requirements (Zhuang et al., 2024).

Technically, AI agents are built upon modular architectures. OpenAI identifies four core components—planning, execution, tools, and memory (OpenAI, 2023)— while others emphasize perception, control, and action (Xi et al., 2023). Within such architectures, AI agent‐based MASs continuously optimize performance through information exchange and reflective iteration.

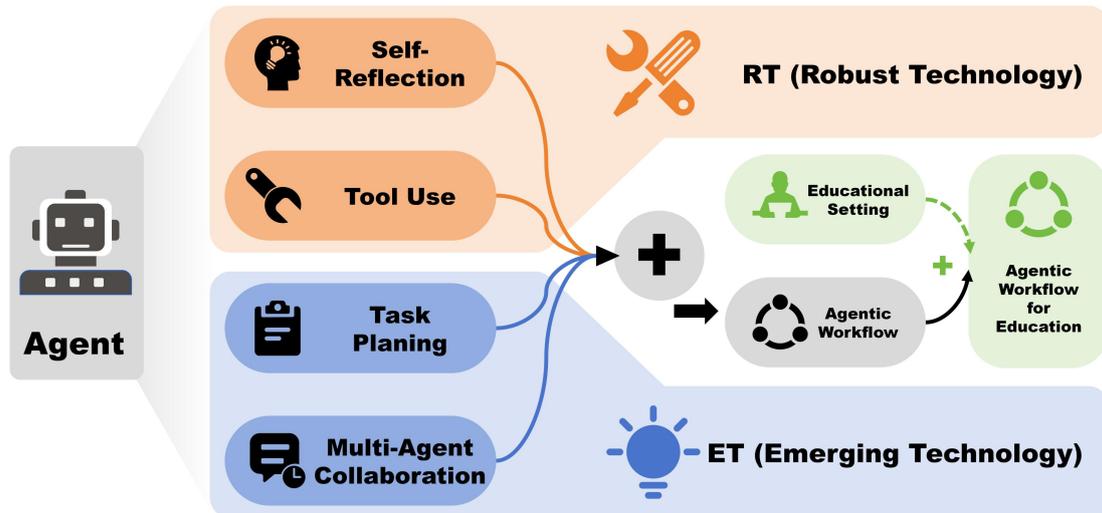

*Figure 2*. The agentic workflow for education and its structural components

This study builds on our earlier work introducing the von Neumann MAS Framework (vNMF) (Jiang, Li, Zhou, et al., 2024), rooted in classical computer architecture (Zhang et al., 2020). vNMF emphasizes processor‐controller decoupling, stored-program adaptability, and input/output interfacing to support agent interaction with educational environments. It decomposes MASs into four components―processor, memory, controller, and I/O devices―mapped to task decomposition, self-reflection, memory processing, and tool invocation (Jiang, Li, Zhou, et al., 2024).

Figure 2 presents the theoretical structure of AWE, aligning four agent capabilities―self-reflection, tool invocation, task planning, and multi-agent collaboration―with two tiers of technological maturity. While the former two are well established, the latter two remain emerging frontiers. Through capability fusion and contextual adaptation, the proposed AWE framework enables intelligent workflows for applications such as exercise generation, automated grading, and AI-assisted tutoring, offering scalable and adaptable support for students, educators, and learning systems.

## 3. Paradigm Shift: From Linear Interaction to Nonlinear Workflows

*3.1 Human-Computer Interaction: From Linear Text Exchanges to Nonlinear Instantiation*

The AWE represents a paradigm shift in how educational systems interact with LLMs, advancing innovation across three core dimensions (see Figure 3). Traditional AI systems based on LLMs rely heavily on a linear "prompt‐response" interaction pattern, which imposes significant constraints (A. Singh et al., 2024). In educational contexts, while linear question-answering based on LLMs has achieved partial automation, users must continuously provide new inputs to advance the task, resulting in inefficiencies and difficulty handling complex, multi-step tasks.

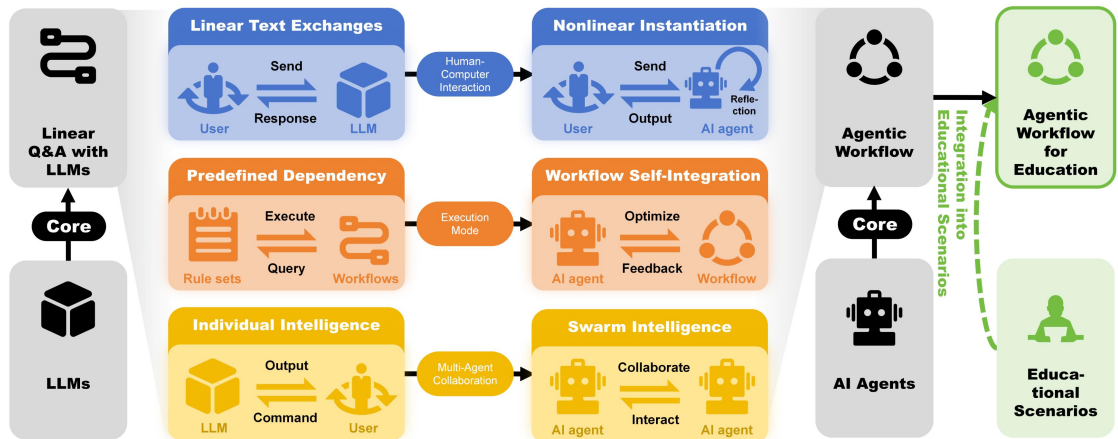

*Figure 3.* Paradigm shift in the agentic workflow for education

In contrast, AWEs leverage AI agent‐based nonlinear workflows during instantiation by integrating capabilities such as self-reflection, tool invocation, task planning, and multi-agent collaboration. These workflows enable agents to autonomously decompose, execute, and optimize tasks, thereby improving the quality and coherence of outcomes (Guan et al., 2024; X. Li et al., 2024). For example, a teacher can configure multiple AI agents to handle sub-tasks including domain knowledge integration, quiz generation, automated problem solving, and distractor design — automatically generating assessment materials without the need for manually refining instructional prompts.

*3.2 Execution Mode: From Predefined Dependency to Workflow Self-Integration*

The AWE framework marks a transition from manually predefined workflows to systems capable of dynamic self-integration. Traditional large-model-driven approaches require educators to design and orchestrate workflows explicitly — an effort-intensive and rigid process ill-suited to dynamic educational environments. In contrast, educational AI agents can autonomously write executable code and adjust workflows during runtime. For instance, based on real-time learning analytics and student feedback, AI agents can generate personalized learning trajectories and embed relevant instructional resources and assessment tools — all without human intervention (X. Li, 2024; Ye et al., 2023).

*3.3 Multi-Agent Collaboration: From Individual Intelligence to Swarm Intelligence*

With the adoption of MASs, Swarm Intelligence (SI) has emerged as a key capability for addressing complex educational scenarios. MASs consist of multiple autonomous AI agents that communicate and collaborate with one another, demonstrating superior performance and adaptability compared to single-agent systems (Chen et al., 2024; Wang et al., 2024). This self-organizing distributed intelligence accounts for many of the practical advantages of MASs in educational use cases (Y. Wu et al., 2024).

For instance, in personalized recommendation systems, MASs leveraging collaborative filtering and multi-dimensional association rules can achieve highly accurate learning resource matching (Khaing & Kalayar, 2024). In homework assessment, agents can divide tasks such as review, feedback generation, and concept mapping, significantly improving both efficiency and instructional quality (R. Wei et al., 2024). Domain expert AI agents that incorporate technologies such as retrieval-augmented generation not only provide high-quality instructional support but also help detect and correct hallucinations generated by LLMs (Shi et al., 2023).

## 4. Application Pathways of the Agentic Workflow for Education

The AWE provides a formalized representation of educational tasks, enabling AI agents to autonomously engage in instructional processes, participate in teaching management, and enhance the overall quality of education. This study categorizes AWE applications into four primary domains based on the scenarios, functions, and feature dimensions of AI agent participation in educational transformation: (1) constructing integrated learning environments, (2) supporting personalized AI-assisted learning, (3) simulating learning scenarios in risk-free settings, and (4) enabling precise educational decision-making.

### *4.1 Constructing Integrated Learning Environments*

In the context of complex problem-solving, AWE facilitates collaborative task execution through MASs. Tasks are decomposed into multiple functional modules, and coordinated interactions among AI agents result in the autonomous formation of integrated solutions. In educational settings, integrated intelligent learning environments typically require the coordination of multiple agents to address complex instructional needs, such as online platform design or virtual learning environments. AI agents enhance automation and personalization in such environments.

For example, a virtual classroom sandbox was developed using MASs to integrate functions such as lesson planning and teacher–student Q&A, supporting instructional preparation (Shi et al., 2023). An AI agent framework based on LLMs, consisting of modules for task definition, planning, and instructional capabilities, was proposed to enable agent collaboration, communication, and competition (Jiang et al., 2025). Similarly, virtual teaching assistant agents were developed to foster collaborative learning environments and improve students' knowledge construction and academic performance (R. Wei et al., 2024).

### *4.2 Supporting Personalized AI-Assisted Learning*

Through iterative self-feedback, AWE enhances agents' perception and expressive capacities, enabling them to generate high-quality, personalized instructional support.

In practice, AI agents can provide targeted feedback to learners, recommend suitable learning strategies, and promote adaptive learning environments.

On one hand, AI agents support learners by analyzing collaborative learning patterns and providing timely feedback that improves engagement and learning efficiency (Bruner, 2023; Martens, 2023). On the other hand, AI agents can dynamically adopt roles such as teacher, peer, or advisor to engage more deeply in the learning process. For example, the introduction of AI agents into group learning environments was shown to enhance participation and facilitate creative idea generation through the processing of unstructured information (Cvetkovic et al., 2023). Role-switching capabilities have also become a new design dimension for AI agents; reinforcement learning can be used to train agents in behavior-switching strategies to determine optimal moments for role transitions. Beyond short-term assistance, AI agents can offer long-term planning and decision support, helping learners pursue sustained educational goals (Nazir et al., 2023). These developments may be further enriched by wearable sensor data, which provide insights into learners' engagement and cognitive states (Hong et al., 2025).

### *4.3 Simulating Learning Scenarios in Risk-Free Settings*

Simulated learning environments represent a key application of AWE in instructional research. By using AI agents to simulate learner behavior across varied scenarios, educators can evaluate the potential effects of interventions or identify hidden instructional issues before real-world implementation.

For instance, simulated environments can be used to model classroom behaviors and assess their consequences, offering early warnings and pedagogical insights. One study simulated school environments to identify external and micro-level factors that negatively affected virtual students (Lee et al., 2022; Zheng et al., 2022). Another study used classroom assessment data and agent-based modeling to examine how attention-related factors interact and influence learning outcomes, helping teachers address issues such as low student attention (Alharbi et al., 2021).

Furthermore, agent-based simulations can replicate a range of student behaviors to test interventions with minimal risk. For example, the EduAgent framework combines cognitive priors with large models to reason about behavioral relationships and simulate authentic student learning processes, supporting data-driven instructional interventions (Xu et al., 2024).

### *4.4 Enabling Precise Educational Decision-Making*

Data mining is a core capability of AWE, allowing AI agents to analyze large volumes of educational data using distributed frameworks, supplement missing data with generative tools, and provide actionable insights for teachers, students, and administrators.

On one hand, agents can recommend personalized resources and strategies. For example, a recommendation agent built on LLMs provides interactive and real-time learning suggestions aligned with students' interests (Huang et al., 2025).

Other systems use AI agents to manage the allocation of educational videos and digital bandwidth, thereby improving the quality of learners' digital experiences (Molnar et al., 2023). On the other hand, AI agents support predictive analytics and strategy generation. The AutoGen framework allows developers to construct agent-based applications through multi-agent dialogue, enabling flexible and personalized learning paths and the automatic generation of instructional materials (Q. Wu et al., 2023). Distributed data mining based on MASs has also been found effective in predicting student performance and generating instructional strategies (Nazir et al., 2023).

## 5. Conclusion

This study proposes the agentic workflow for education as a novel paradigm that integrates AI agent autonomy, multi-agent collaboration, and task-driven workflows to address complex educational needs. By bridging theoretical insights and practical implementations, AWE demonstrates the potential to enhance learning experiences, support large-scale instructional tasks, and drive pedagogical innovation across diverse educational contexts.

## Acknowledgements

This work was jointly supported in part by the Fundamental Research Funds for the Central Universities, and the ECNU Academic Innovation Promotion Program for Excellent Doctoral Students (both under Grant YBNLTS2025-008). This work was also partially supported by the National Natural Science Foundation of China, under the Grant 62477012, and the Natural Science Foundation of Shanghai, under the Grant 23ZR1418500, and the Special Foundation for Interdisciplinary Talent Training in "AI Empowered Psychology / Education" of the School of Computer Science and Technology, East China Normal University, under the Grant 2024JCRC-03.